%% file: NeutrinoDirectionality.tex
\documentclass[reprint,superscriptaddress,aps,prd]{revtex4-2}
\usepackage{amsmath,amssymb}
\usepackage{svrsymbols}
\usepackage{graphicx}
\usepackage{times}
\usepackage[T1]{fontenc}
\usepackage{xcolor}
\usepackage{lineno}
\usepackage{bm}
\usepackage{textcomp, gensymb}
\usepackage{hyperref}
\usepackage{tabularx}
\usepackage{adjustbox}
\usepackage{multirow}
\usepackage{microtype}
\usepackage{svg}
\usepackage[separate-uncertainty=true,multi-part-units=single]{siunitx}
\usepackage[all]{hypcap}

\hypersetup{
   colorlinks   = true, 
   urlcolor     = blue, 
   linkcolor    = blue, 
   citecolor   = blue 
}

\newcommand{\NIST}{National Institute of Standards and Technology (NIST) \renewcommand{\NIST}{NIST}}
\newcommand{\NBSR}{National Bureau of Standards Reactor (NBSR) \renewcommand{\NBSR}{NBSR}}
\newcommand{\INL}{Idaho National Laboratory (INL) \renewcommand{\INL}{INL}}
\newcommand{\ATR}{Advanced Test Reactor (ATR) \renewcommand{\ATR}{ATR}}
\newcommand{\ORNL}{Oak Ridge National Laboratory (ORNL) \renewcommand{\ORNL}{ORNL}}
\newcommand{\HFIR}{High Flux Isotope Reactor (HFIR) \renewcommand{\HFIR}{HFIR}}
\newcommand{\LLNL}{Lawrence Livermore National Laboratory (LLNL) \renewcommand{\LLNL}{LLNL}}

{

\begin{document}


\title{Reactor Antineutrino Directionality Measurement with the PROSPECT-I Detector}

\input{AuthorList2022.tex}

\collaboration{The PROSPECT Collaboration}
\homepage{http://prospect.yale.edu \\ email: prospect.collaboration@gmail.com}
\noaffiliation


\begin{abstract}
  The PROSPECT-I detector has several features that enable measurement of the direction of a compact neutrino source. In this paper, a detailed report on the directional measurements made on electron antineutrinos emitted from the High Flux Isotope Reactor is presented. With an estimated true neutrino (reactor to detector) direction of $\phi = \ang{40.8} \pm \ang{0.7}$ and $\theta = \ang{98.6} \pm \ang{0.4}$, the PROSPECT-I detector is able to reconstruct an average neutrino direction of $\phi = \ang{39.4} \pm \ang{2.9}$ and $\theta = \ang{97.6} \pm \ang{1.6}$. This measurement is made with approximately \num{48000} Inverse Beta Decay signal events and is the most precise directional reconstruction of reactor antineutrinos to date.
\end{abstract}

\maketitle

\section{Introduction}

Directional neutrino measurements have been valuable to investigations of atmospheric neutrino oscillations~\cite{SuperK} and in neutrino astronomy~\cite{IceCube}. The direction of high-energy neutrinos in these experiments is reconstructed by the timing and spatial distribution of Cherenkov photons from recoiling electrons. However, for low energy (\unit{\mega \eV}-scale) neutrinos, directional reconstructions are challenging due to the short ranges of the products of neutrino interactions relative to detector position resolution. The technique demonstrated here for the directional measurement of low energy electron antineutrinos may be beneficial for a number of applications, including reactor monitoring and nuclear non-proliferation, supernovae pointing, and better understanding of geo-antineutrino (geo-$\antineutrino_e$) source distributions.

Low-energy electron antineutrinos (such as those produced in nuclear reactors) are typically detected through Inverse Beta Decay (IBD) on hydrogen in organic or aqueous media, resulting in a neutron and a positron:
\begin{equation}
  \antineutrino_e + p \rightarrow n + \positron
\end{equation}

\noindent The positron annihilates within \qty{1}{\nano \second}, while the neutron is captured with an average time of \qtyrange{50}{200}{\micro \second}, depending on the presence of neutron capture additives such as dissolved metals with large thermal neutron cross-sections. The resulting distinctive coincidence signal allows for isolation of IBD interactions from background events.

Conservation of energy and momentum requires that the average outgoing neutron direction is strongly downstream from the reactor, as described in detail by Vogel and Beacom \cite{Vogel}. However, scattering of the neutron during thermalization and subsequent neutron diffusion prior to capture substantially degrades any directional information. The mean neutron displacement in liquid scintillator is approximately \qty{1.6}{\centi \metre}, while the spread due to scattering and diffusion is about \qty{6}{\centi \metre}. Thus, the angular resolution for a single antineutrino is necessarily very poor, even before finite detector position resolution is considered. Nevertheless, when there are many events from the same source, a statistical reconstruction of antineutrino direction is possible by averaging the positron-neutron capture displacement.

Simulation studies have been used to investigate the directional measurement ability of the IBD interaction in a wide range of applications and detection implementations \cite{Beacom:1998fj,CHOOZ,Domogatsky:2004be,Batygov:2005zz,Hochmuth:2006gz,Shimizu:2007zzc,Hochmuth:2007gv,Jocher:2013gta,Pac:2013apo,Dawson:2014lga,Geoneutrinos,Safdi:2014hwa,Fischer:2015oma,mTC:2016yys,AIT-WATCHMAN:2019gnv,Seo:2020zmc,Seo:2020fdu,Shin:2020yic,DirectionalPreSupernova,PreSupernovaLiLS,Sutanto:2021xpo,Radermacher:2022pnt,Duvall:2024cae}. Previous segmented detector experiments have observed the expected displacement with reactor antineutrinos \cite{Zacek,Boehm2000}. Monolithic detectors such as CHOOZ and Double Chooz have also performed measurements of the incident antineutrino direction, utilizing Gd as a neutron capture agent \cite{CHOOZ, Caden}. Different experiments have used other capture isotopes. The spatial extent of the detected reaction products from neutron capture on ${}^6{\rm Li}$ via ${}^6{\rm Li}(n,\alpha)^{3}{\rm H}$ is greatly reduced in comparison to that of gamma production from neutron capture on Gd or H nuclei, providing benefits in terms of position resolution \cite{Abbes:1995nc,Shimizu:2007zzc,KIFF2011412,PROSPECT:2013phf,Geoneutrinos,NuLat:2015wgu,Ryder:2015sma,Haghighat:2018mve,Li:2019sof}, which could in turn provide improved directional measurements \cite{Shimizu:2007zzc,Geoneutrinos,Sutanto:2021xpo,DirectionalPreSupernova}. This paper presents a reconstruction of the IBD antineutrino direction with the ${}^6{\rm Li}$-doped liquid scintillator detector of the PROSPECT experiment, improving on previous reactor antineutrino directionality measurements. This reconstruction is of the mean antineutrino direction, and not a measurement of angular resolution.
\section{Motivation}

Reactor antineutrino detection has the potential to be a new tool for nuclear safeguards \cite{Bernstein2019}. With sufficient counting statistics, information on reactor power and fissile content can be obtained in a non-intrusive way \cite{ReactorMonitoring, Klimov1994, Bernstein2008}. Sensitive detectors could also allow for distinguishing fuel enrichment as well as continuous monitoring even while the reactor is shut down, provided low enough background \cite{Christensen2014, An2017}. Antineutrino directional information has been suggested as a way to monitor individual reactors at multi-unit facilities \cite{Akindele2021} or to suppress background from cosmogenic sources and other distant reactors in far-field monitoring proposals \cite{Hellfeld2017, AIT-WATCHMAN:2019gnv}. Detectors designed to be sensitive to reactor IBD energies can employ similar directional methods to those presented in this paper to enhance their applications in reactor monitoring.

Directional detection capabilities are also attractive for multiple physics measurements. Liquid scintillator-based detectors are well-suited to determine the direction of a supernova using the IBD interaction \cite{Beacom:1998fj,CHOOZ}. The measurement in this paper demonstrates the concept of such a measurement. Other methods to determine supernova direction rely on reconstruction of the outgoing products of neutrino charged-current with nuclei \cite{Abi2021,Abe2016}. It is expected that neutrinos are also emitted before the onset of core collapse in supernovae~\cite{PreSNSensitivity}. Though these pre-supernova neutrinos are lower energy and luminosity than those emitted during collapse, current models estimate that they could be detected up to a day before the explosion, if not earlier \cite{Presupernova}. A directional detection of pre-supernova neutrinos could enhance SNEWS (SuperNova Early Warning System), a network of neutrino detectors with the goal of providing early warnings to astronomers so that telescopes can be pointed towards supernovae \cite{SNEWS}. It is suggested that IBD events detected by large ($\mathcal{O}$(\qty{10}{\kilo \tonne} mass)) detectors could offer the best potential for localizing pre-supernova neutrinos \cite{DirectionalPreSupernova, PreSupernovaLiLS}. In addition, directional reconstruction of low-energy (< \qty{1}{\giga \eV}) atmospheric neutrinos in large liquid scintillator detectors would enable access to oscillation phenomenology sensitive to the leptonic CP-violating phase \cite{DUNECPV}.

Since approximately \qty{50}{\percent} of the Earth's internal heat is radiogenic, neutrino detectors have already been able to estimate the radiogenic heat using IBD events caused by geo-$\antineutrino_e$ interactions to provide insight about the inner workings of our planet \cite{RadiogenicHeat}. However, these detectors have not been able to access geo-$\antineutrino_e$ directional information due to low event rates and use of undoped liquid scintillators. Directional measurements could allow for model independent information about the spatial distribution of $\antineutrino_e$ sources in the Earth's crust and mantle \cite{Leyton2017,Geoneutrinos,Domogatsky:2004be,Batygov:2005zz,Hochmuth:2006gz}.

\section{PROSPECT}

\begin{figure}[ht]
  \centering
  \includegraphics[width=0.48\textwidth]{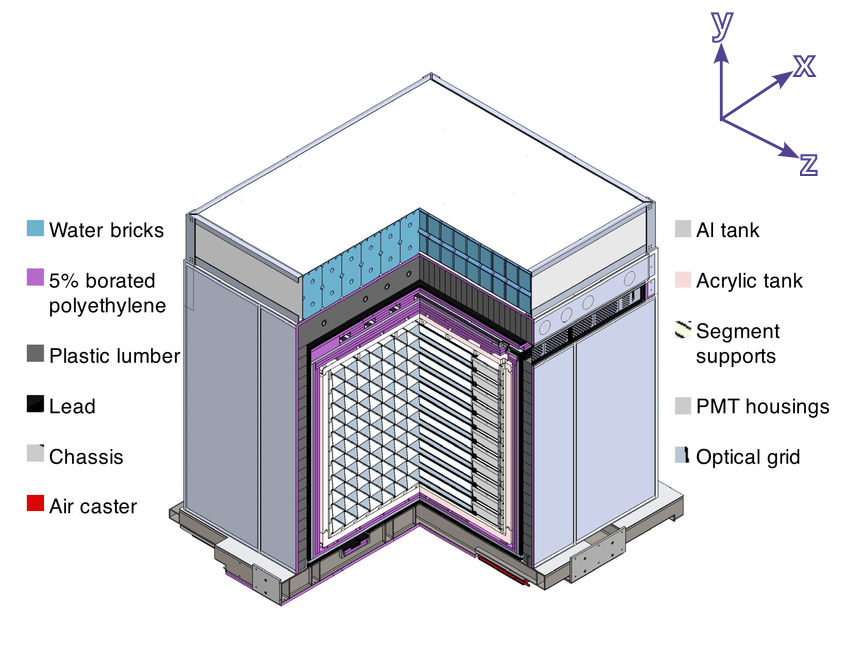}
  \caption{A cutaway of the PROSPECT-I antineutrino detector. The reactor is located towards the negative $x$ and $y$ axes, and positive $z$ axis.}
  \label{fig:PROSPECT}
\end{figure}

The Precision Reactor Oscillation and Spectrum Experiment (PROSPECT \cite{PROSPECT_PRD}) is a short-baseline reactor antineutrino experiment at Oak Ridge National Laboratory. The PROSPECT-I detector was deployed to search for eV-scale sterile neutrino oscillations and to make a precise measurement of the energy spectrum of antineutrinos emitted from the \HFIR\ at baselines from \qtyrange{7}{9}{\meter}. PROSPECT measurements have tested our understanding of the standard model of particle physics, deepened our understanding of nuclear processes in a reactor, and achieved the highest signal-to-background ratio for a surface-based neutrino detector.

The PROSPECT-I detector \cite{PROSPECT_PRD} was a \qtyproduct{2.0 x 1.6 x 1.2}{\metre} rectangular acrylic shell containing about four tons of liquid scintillator loaded with \textsuperscript{6}Li (LiLS) \cite{LiLS}. Thin, specularly reflecting panels divide the LiLS volume into an array of \num{154} optically isolated rectangular segments of \num{14} segments in $x$ and \num{11} segments in $y$. The segments are \qtyproduct{14.6 x 14.6 x 117.6}{\centi \metre} each with the long axes along $z$, viewed on both ends by \qty{12.7}{\centi \metre} PhotoMultiplier Tubes (PMTs) \cite{Reflectors}. These optical segments are each tilted at \ang{5.5} from the vertical axis. The segmented design of PROSPECT and its double-ended PMT readout enable good position reconstruction. The detector operated with minimal overburden from the \HFIR\ building. Figure \ref{fig:PROSPECT} shows a diagram of PROSPECT-I.

\begin{figure*}[ht!]
  \includegraphics[width=0.48\textwidth]{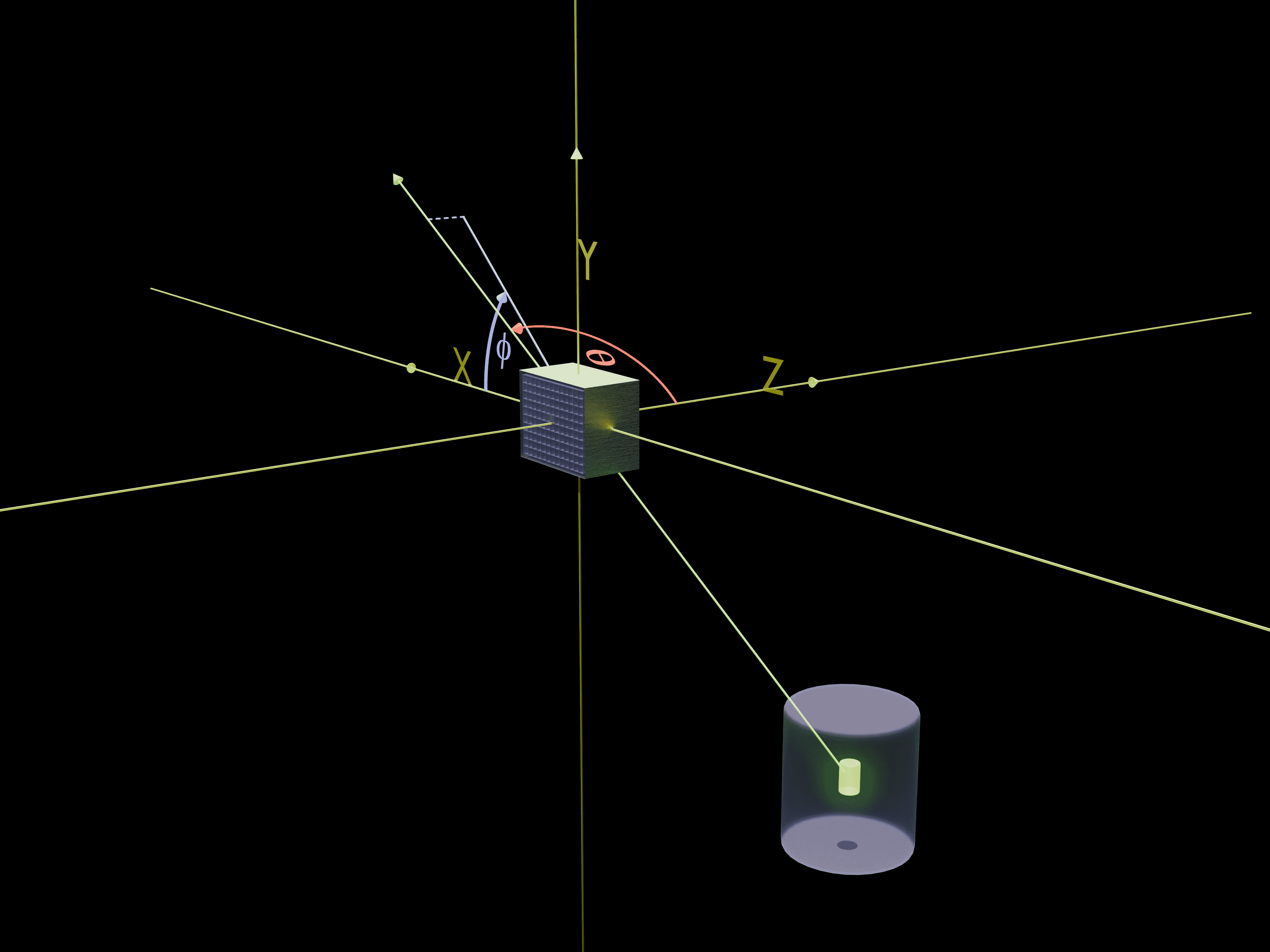}
  \hfill
  \includegraphics[width=0.48\textwidth]{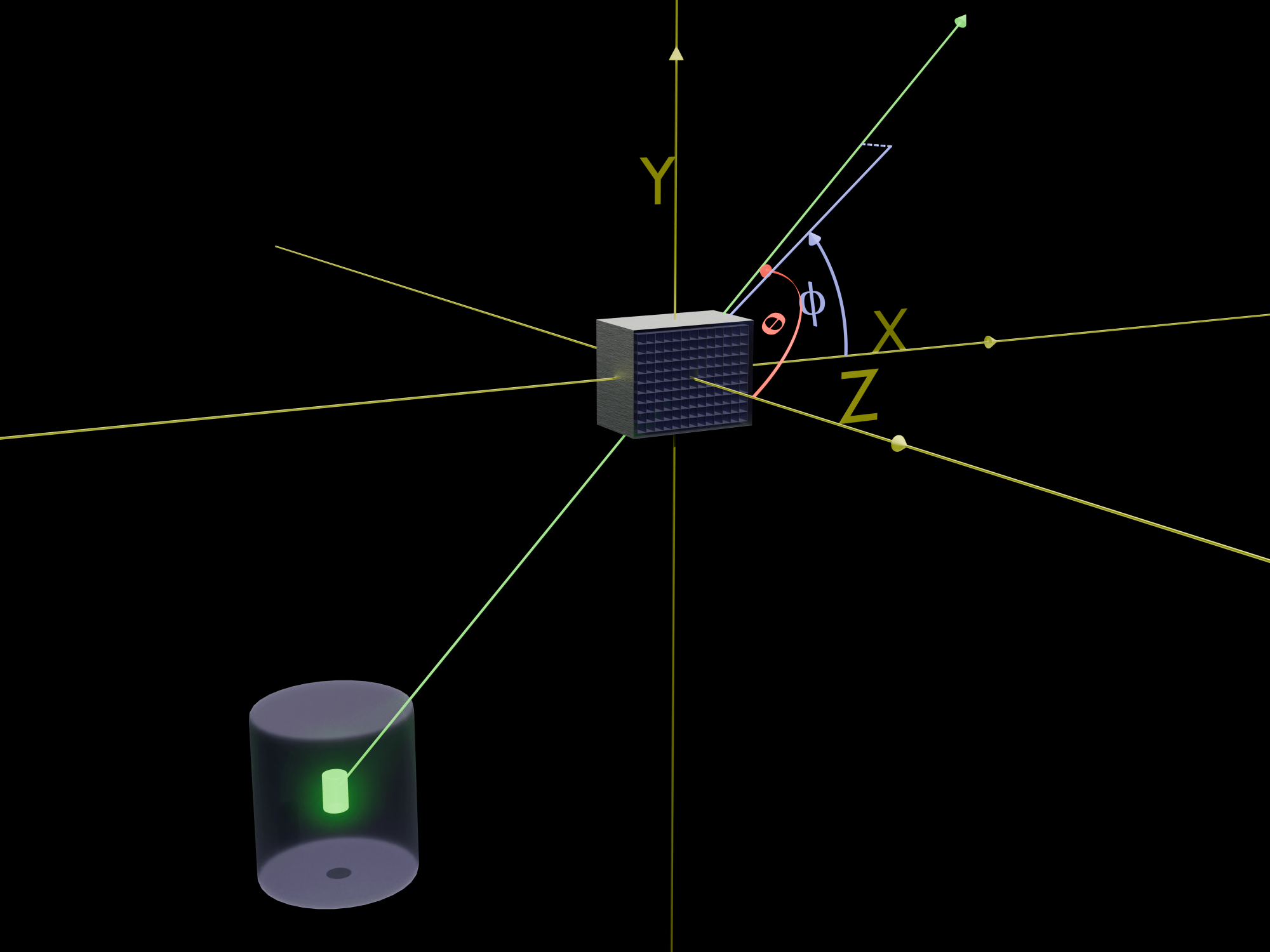}
  \caption{A 3D rendering of the PROSPECT-I detector (placed at the origin), and the \HFIR\ reactor core (glowing green) encased in the pressure vessel (blue). The positive $z$ direction is parallel to the long axis of the segment, slightly towards the reactor. The positive $y$ axis is vertical. The positive $x$ direction is away from the reactor.  The expected average antineutrino direction, pointing away from the center of the reactor core towards the PROSPECT-I detector, is shown as a green arrow. The azimuthal and zenith angles being measured are shown, with $\phi$ in light blue and $\theta$ in red. $\phi$ is the angle that the antineutrino vector projects on the $xy$ plane, from the $x$ axis towards the $y$ axis. The projection vector is shown by the solid and dashed light blue lines. $\theta$ is the angle from the $z$ axis towards the antineutrino vector.}
  \label{fig:prospect_reactor_xyz}
\end{figure*}

The $^6$Li is used because it has both a high neutron capture cross-section and facilitates the identification of the neutron capture using Pulse Shape Discrimination (PSD). PSD is utilized to identify the type of particle interaction occuring in the detector, taking advantage of the fact that liquid scintillator response time distributions vary according to particle interaction types. Previous measurements of incident reactor antineutrino direction performed by Chooz \cite{CHOOZ} and Double Chooz \cite{Caden} utilized monolithic Gd-loaded liquid scintillator detectors. Gd produces high-energy, diffuse neutron capture gamma rays while $^6$Li emits a triton and alpha after neutron capture, which lead to a highly localized neutron capture signal (as well as a dense ionization cloud, enabling PSD) \cite{Geoneutrinos}. By doping at a mass fraction of \qty{0.082 \pm 0.001}{\percent},  PROSPECT-I achieves an average capture time of \qty{50}{\micro \second} \cite{LiLS}.

The analysis described in this paper uses data taken from March 5 to October 6 of 2018. A significant challenge of the PROSPECT-I dataset is the varying number of operating PMTs during the data taking period. Intrusion of LiLS into the PMT housings caused current instability in some PMT voltage dividers. This led to a subset of PMTs being rendered inoperative throughout the data taking period considered here. The data taking has been split into five periods corresponding to the five \HFIR\ fuel cycles, each with a different configuration of live segments and inoperative segments. Segments with only one PMT operational are now used to improve background rejection with SEER (Single Ended Event Reconstruction), increasing PROSPECT-I's already high signal-to-background ratio. "Data splitting" and SEER are the same techniques used in the final PROSPECT-I spectrum measurement \cite{PROSPECT_PRL}.

\section{Methods}
\label{sec:methods}
\subsection{Event Reconstruction and IBD Selection}

For each PROSPECT IBD candidate, the position, energy, and PSD are reconstructed from the PMT waveforms \cite{PROSPECT_PRD}. IBD positron event candidates are denoted as prompt events and neutron capture candidates as delayed events. The $x$ and $y$ positions of prompt and delayed events are reconstructed as the midpoint of the optical segment. The $z$ positions are reconstructed using relative pulse timing and amplitude from the PMTs at either end of the segment. The segment with the largest energy deposit is assumed to correspond to the IBD vertex. A PSD value is calculated for each recorded waveform using the tail-over-total ratio, optimized to maximize neutron capture discrimination. Features from individual PMTs are grouped into multi-segment clusters within a \qty{20}{\nano \second} arrival time window. Reconstructed pulses are then defined as paired waveforms of PMTs that are stored and analyzed together. The reconstructed energy ($E_{rec}$), is defined as the sum of reconstructed energies of all contained pulses within an event.

The initial selection of IBD candidate coincident pairs proceeds as described in Reference \cite{PROSPECT_PRL}. The prompt event must have an $E_{rec}$ between \qtyrange{0.8}{7.4}{\mega \eV} and individual reconstructed pulse PSD values within \num{2.0} standard deviations of the calibrated electron-like PSD mean. The delayed event must be a single-segment event with an ($E_{rec}$, PSD) pair that is consistent with $n$-Li capture. This delayed event is required to occur within \qty{120}{\micro \second} of the prompt event and its segment must be the same as or vertically/horizontally adjacent to that of the prompt event. If the prompt and delayed events occur in the same segment, their difference in $z$ is required to be less than \qty{140}{\milli \metre}; if the segments are adjacent, the difference in $z$ must be less than \qty{60}{\milli \metre}. IBD candidates spread across diagonal segments are discarded due to the low event rate.

IBD candidates are selected during both reactor-on and reactor-off periods. For each, accidental coincidences are subtracted, where the accidental rate is estimated with an offset time window that is ten times longer than the IBD coincidence window. Finally, for all parameters of interest, the reactor-off background, scaled by reactor live-time, is subtracted from the reactor-on data to give the IBD signal. \num{61029} IBD event-pairs were collected over \num{95.62} (\num{72.69}) reactor-on (off) days. This IBD event-pair sample is used to reconstruct the incident antineutrino direction.

\subsection{Directional Reconstruction}

The relative positions of PROSPECT-I and the \HFIR\ core as well as the spherical angles being measured in this analysis are shown in Figure \ref{fig:prospect_reactor_xyz}. Reconstructing the positron annihilation and neutron capture positions are the two key elements for measuring the direction of the incoming antineutrinos. The Vogel-Beacom model predicts that IBD positron displacement is biased slightly upstream to the reactor, while neutrons are expected to move downstream from the reactor \cite{Vogel}. In simulations of PROSPECT-I, described in Section \ref{subsec:sims}, the average positron displacement from the IBD interaction point was found to be $\sim$\qty{0.2}{\milli \metre} backwards (towards the reactor), as expected, meaning its mobility can be neglected in the directional determination.

The incident antineutrino direction is determined from the vector, $\bm{s}$, pointing from the reconstructed positron position to the reconstructed neutron capture position. Averaging this vector over $N$ IBD event-pairs gives:

\begin{equation}
  \bm{p} = \frac{1}{N} \sum_{i}^{N} \bm{s}_i = \frac{1}{N} \sum_{i}^{N} (\bm{s}_{d_i} -\bm{s}_{p_i})
  \label{eq:average_vector}
\end{equation}

\noindent where $\bm{s}_{d_i}$ and $\bm{s}_{p_i}$ are the positions of the delayed and prompt events for the $i$-th pair. The vectors $\bm{s}$ should average to a vector $\bm{p}$ that points away from the reactor core. The measurement improves with higher statistics, so PROSPECT-I's full dataset is used for this analysis. The calculation of $\bm{p}$ in the $x$ and $y$ directions, $p_x$ and $p_y$, can be simplified due to the discretization of positions to the segment cross-section midpoint, and the requirement that prompt and delayed events are in the same or adjacent segments. The $x$ and $y$ components of $\bm{s}$ can only be: $-D$, 0, or $+D$, where $D$ is the segment pitch of \qty{14.6}{\centi \metre}. Denoting the background-subtracted number of pairs where $s_\alpha = D$ as $n_{\alpha_+}$, and the number of pairs where $s_\alpha = -D$ as $n_{\alpha_-}$, Equation \ref{eq:average_vector} becomes:

\begin{align}
  p_\alpha = \frac{Dn_{\alpha_+} - Dn_{\alpha_-}}{N} ,
  \label{eq:px}
\end{align}

\noindent for $\alpha \in \{x, y\}$.

The position resolution in $z$ does not have the same discretization. The mean of a Gaussian fit to the background-subtracted $s_z$ distribution determines $p_z$.

\begin{figure*}[ht!]
  \includegraphics[width=0.32\textwidth]{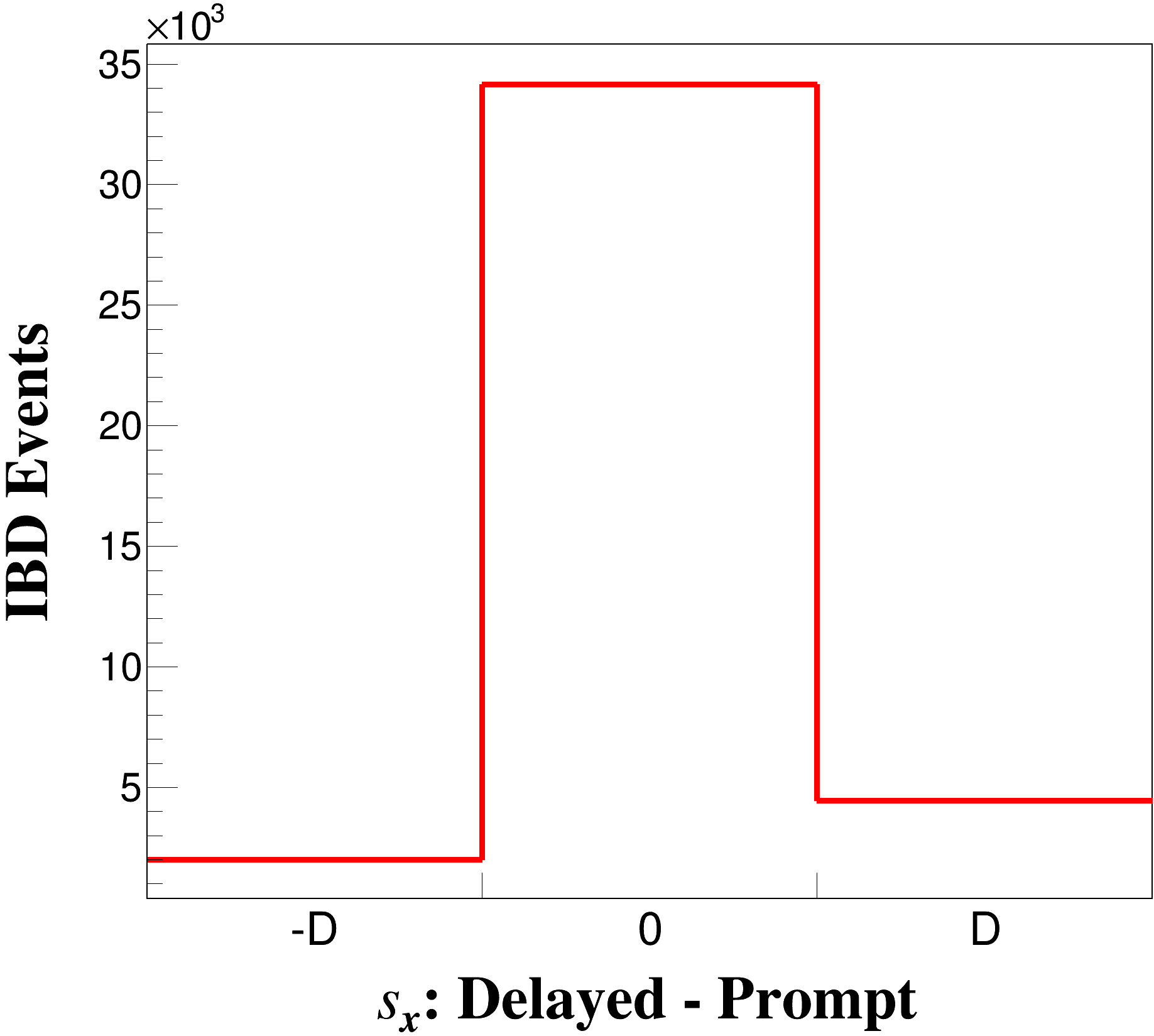}
  \hfill
  \includegraphics[width=0.32\textwidth]{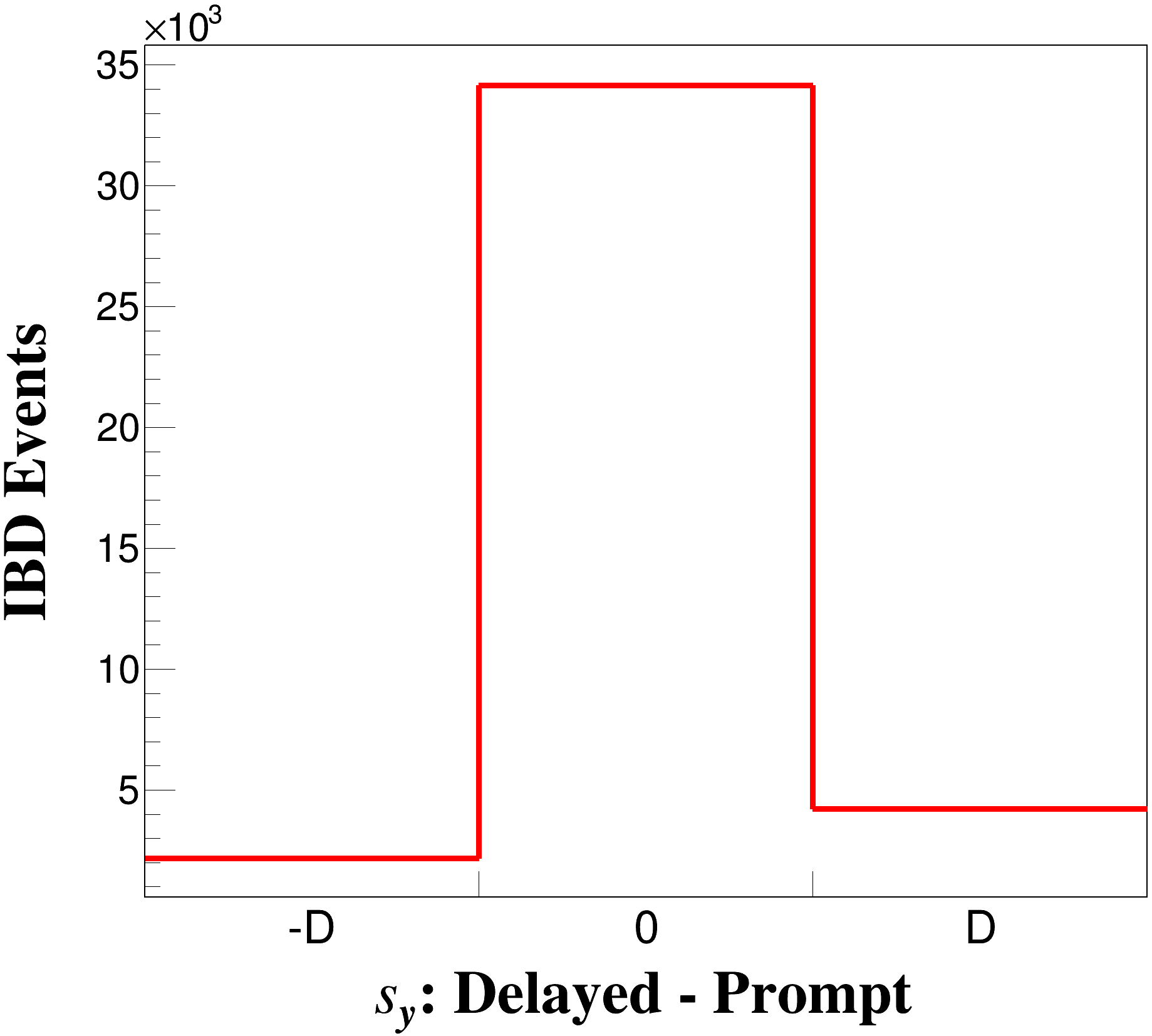}
  \hfill
  \includegraphics[width=0.32\textwidth]{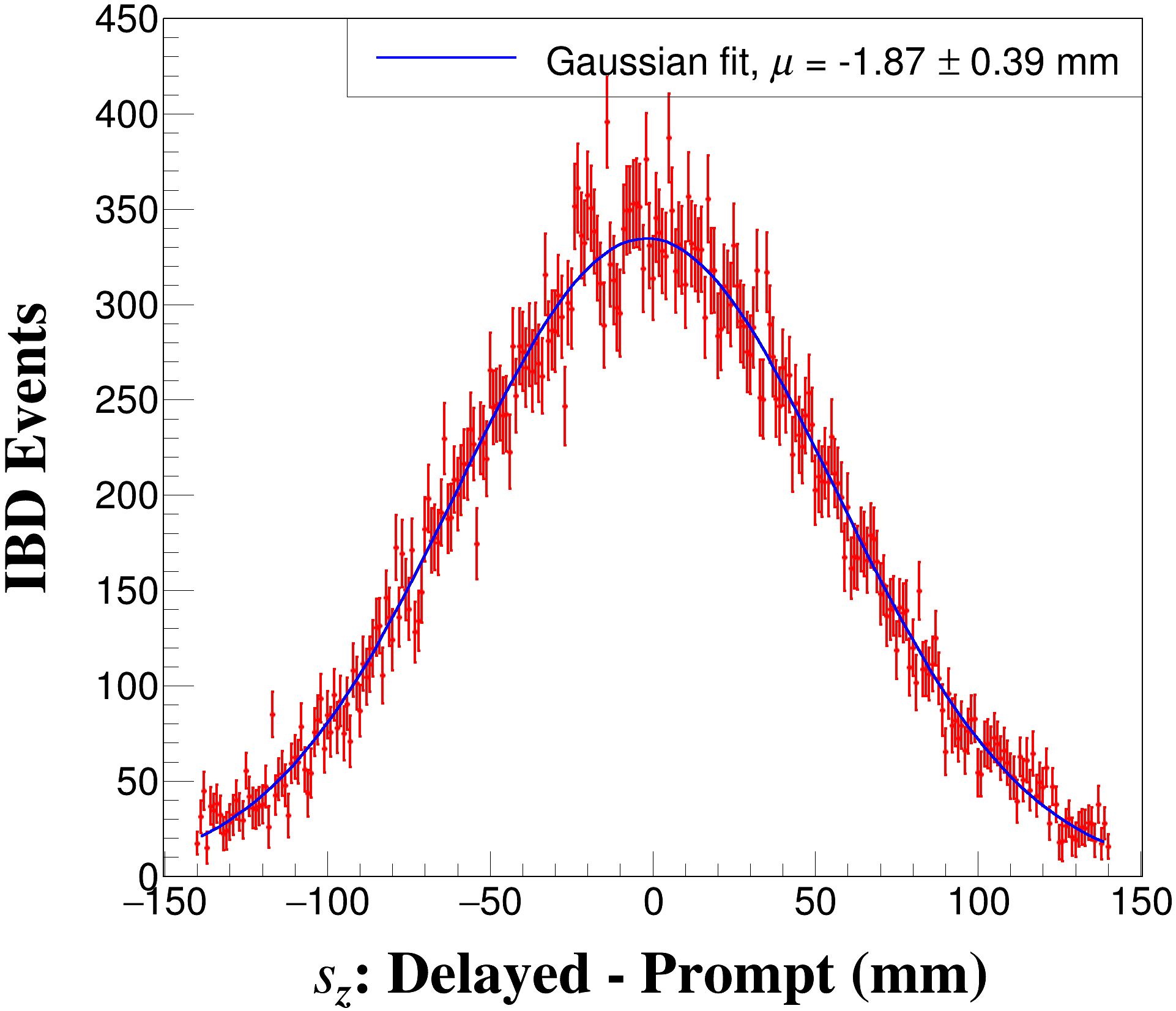}
  \caption{Distributions of the \num{3} components of the vector $\bm{s}$ after background subtraction display the ability of the PROSPECT-I detector to resolve the delayed-prompt position displacement caused by the directionality of the IBD interaction. $D$ refers to the segment width of \qty{14.6}{\centi \metre}. As expected, the majority of prompt and delayed pairs happen in the same segment. The remaining pairs are visibly skewed along the vector that points from the reactor to the detector. The $s_z$ plot showcases the data points in red, with error bars, and the Guassian fit in blue. The gaussian fit determines $\sigma = 58.10 \pm 0.35$ \unit{\milli \metre} with $\chi^2 = 267$ for \num{278} degrees of freedom.}
  \label{fig:xyz distribution}
\end{figure*}

The cuts on $\bm{s}$ from reference \cite{PROSPECT_PRL} were chosen to optimize signal-to-background, but are not ideal for a directional measurement. Specifically, accepting delayed events that are vertically or horizontally adjacent to the prompt event but not both (diagonally adjacent), creates a bias against large displacements in $x$ or $y$. Similarly, the smaller $z$ cut applied to delayed events in adjacent segments creates a selection bias against those events, leading to a too-small determination of the average displacement both transverse to and along the segment axis. In addition, the relatively small \qty{60}{\milli \metre} adjacent-segment $z$ cut removes a substantial part of the tails of the distribution, inhibiting the ability to fit the $s_z$ distribution to precisely determine the mean. To mitigate these effects without expanding the selection cuts, a different set of restricted cuts is applied for the measurement of each component of $\bm{p}$. For the measurement of $p_x$, delayed events in either adjacent segment in $y$ are excluded, and vice versa. Additionally, for the measurements of $p_x$ and $p_y$, the smaller \qty{60}{\milli \metre} $z$ cut is applied equally to both the same and adjacent segments. For the measurement of $p_z$, the fit of the distribution is done only on same-segment events, for which the \qty{140}{\milli \metre} $z$ cut leaves the bulk of the distribution intact. These adjustments remove the potential biases at the cost of statistics. The background-subtracted distribution of events in all three axes can be seen in Figure \ref{fig:xyz distribution}.

\begin{figure}[h]
  \includegraphics[width=0.33\textwidth]{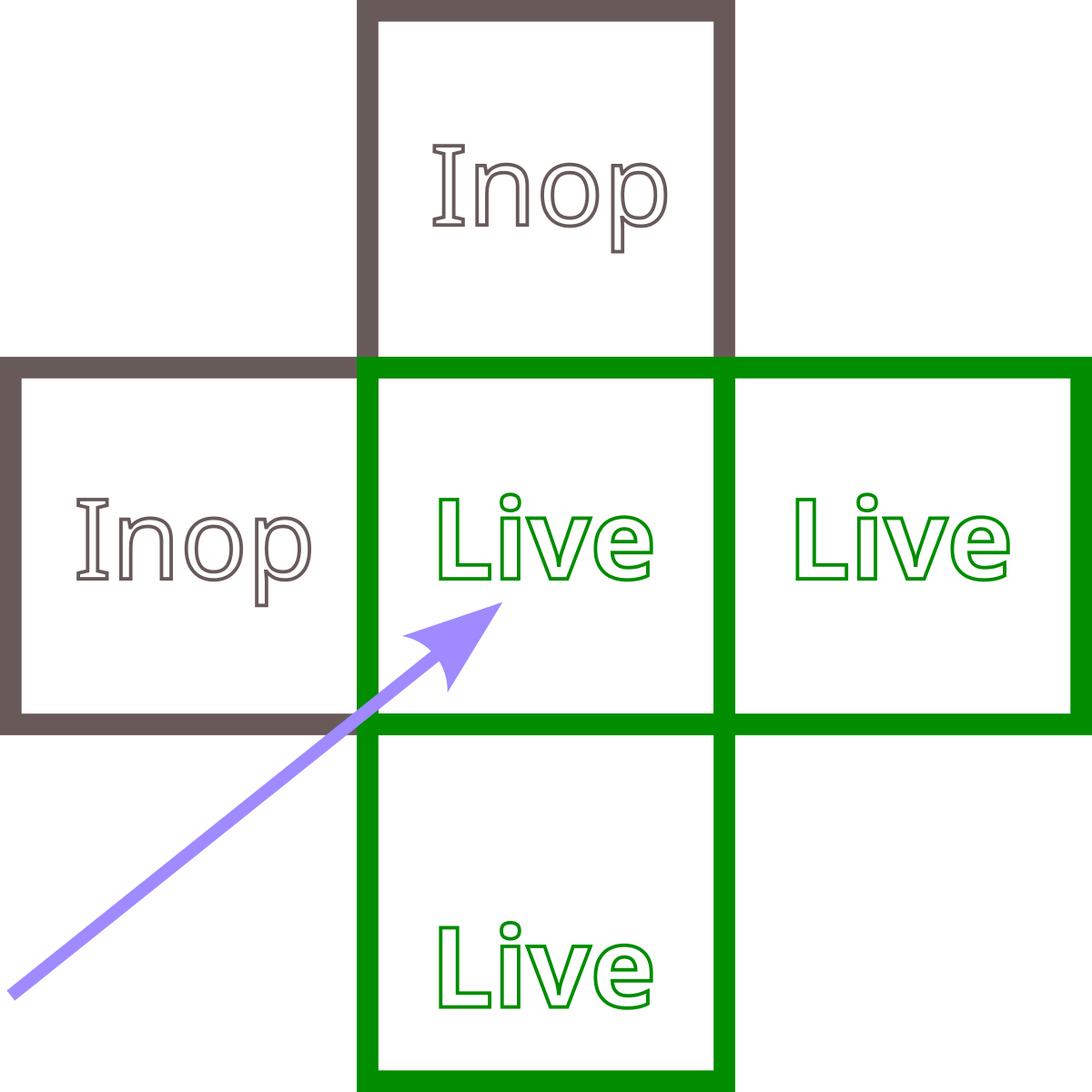}
  \caption{A demonstration of how inoperative segments can cause a bias in the detected antineutrino direction. The blue arrow represents the average antineutrino vector, away from the reactor. Assume the prompt event is in the center segment. Note how for this particular configuration of live and inoperative segments, the observed average neutron displacement would be down and to the right, instead of up and to the right, biasing the measurement.}
  \label{fig:seg_diagram}
\end{figure}

Another potential source of bias in $p_x$ and $p_y$ is the presence of inoperative segments throughout the detector, as well as edge effects. IBD pairs for which the neutrons are captured in inoperative or non-fiducial segments are missing from the average. Due to the irregular layout of inoperative segments and non-uniform IBD distribution, this can result in a biased average. An example of this effect in shown in Figure \ref{fig:seg_diagram}.

To mitigate the missing neutron problem, a modified analysis is employed to calculate $p_x$ and $p_y$, in which only neutrons moving towards live fiducial segments in each direction are counted. Consider the calculation of $p_x$. Equation \ref{eq:px} can be rewritten as:

\begin{equation}
  p_x = D \frac{n_{x_+} - n_{x_-}}{n_{x_+} + n_{x_-} + n_0} ,
  \label{eq:expanded_px}
\end{equation}

\noindent where $n_0$ is the background-subtracted number of IBD pairs where the prompt and delayed events are in the same segment. Dividing by $n_0$ gives:

\begin{equation}
  p_x = D\frac{r_{+}-r_{-}}{r_{+} + r_{-} + 1} ,
  \label{eq:ratio}
\end{equation}

\noindent where $r_\pm = \frac{n_{x_\pm}}{n_0}$.
\smallskip

\begin{figure}[h]
  \includegraphics[width=0.4\textwidth]{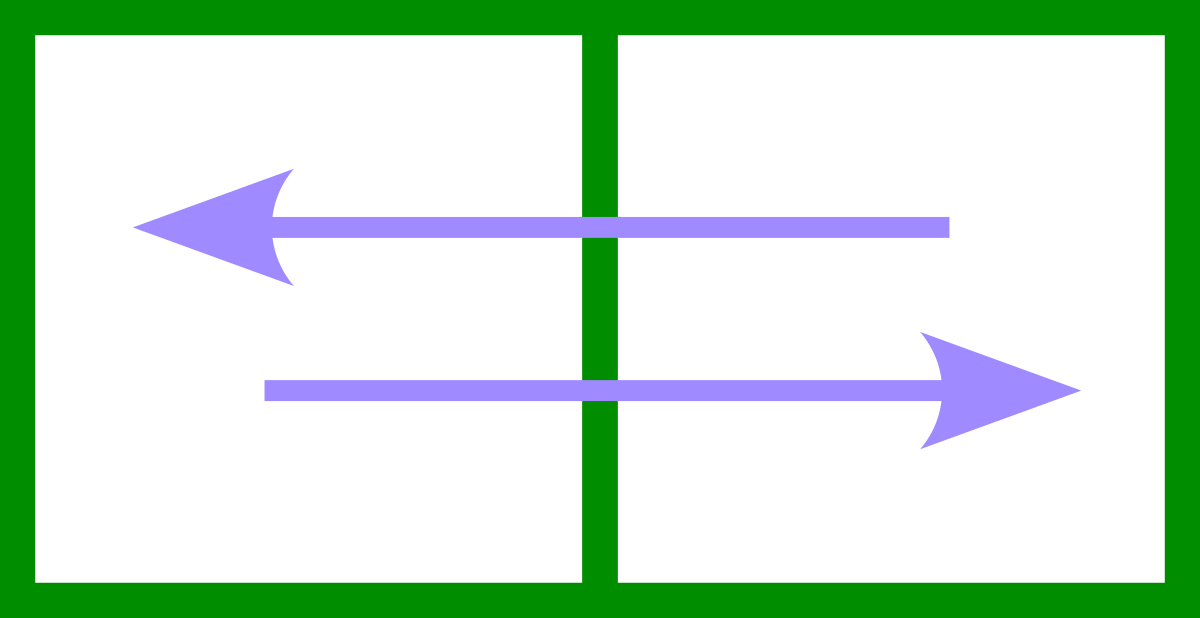}
  \caption{To calculate $p_x$, we can consider only pairs of live segments that are neighbors in the $x$ axis (no lone live segments). Using horizontal pairs, the ratio of delayed events that move to the left and to the right can be calculated, pair by pair. This value can then be averaged over every horizontal pair, giving a measurement of the average neutron displacement in the $x$ direction.}
  \label{fig:seg_arrows}
\end{figure}

The advantage of Equation \ref{eq:ratio} is that $r_+$ and $r_-$ can then be evaluated separately using different event selection criteria, such that the bias from missing neutrons in inoperative segments is avoided. This is achieved by measuring $r_+$ ($r_-$) using only prompt events in segments with a live fiducial segment to the right (left). Effectively, the detector is divided into sub-detectors consisting of adjacent segment pairs, as illustrated in Figure \ref{fig:seg_arrows}. Prompt events in the left segment are used to measure the ratio of neutrons migrating to the right ($r_+$) and prompt events in the right segment are used to measure the ratio of neutrons migrating to the left ($r_-$).

For the implementation of this modified analysis in $x$: first, all lone live fiducial segments are discarded. Only pairs of live segments are considered for the calculation. The existence of live neighbors in $y$ is not relevant to the calculation of $p_x$. Events for which $s_x = \pm D$ are straightforward and treated the same way in both methods. However, events for which $s_x = 0$ can be divided into three categories: $n_{0\pm}$, for which the segment in the positive or negative $x$ direction is the \textit{only} live adjacent fiducial segment in $x$; and $n_{0*}$, for which both adjacent segments in $x$ are live fiducial segments. Note that to evaluate $r_+$ over the whole detector, only pairs where $s_x = D$ or $0$ need to be considered. Similarly, only pairs where $s_x = -D$ or $0$ are necessary to calculate $r_-$. The ratios for the $x$ direction can now be calculated:

\begin{equation}
  r_{\pm}=\frac{n_{\pm}}{n_{0\pm} + n_{0*}},
\end{equation}

\noindent the unbiased ratio of delayed events in the positive or negative $x$ direction to delayed events in the same segment. The optimized ratios calculated here can be utilized in Equation \ref{eq:ratio}.

In the case where $n_{0+}=n_{0-}=0$, as would be the case for an infinite detector with no inoperative segments, $N=n_{+} + n_{-} + n_{0*}$ and Equation~\ref{eq:ratio} reduces to Equation~\ref{eq:px}. But as this is not the case, Equation~\ref{eq:ratio} represents a less biased version of Equation \ref{eq:px} in the presence of inoperative segments and edge effects. A similar procedure is employed to calculate $p_y$ in a less biased way. Results for the $p$-components are shown in Table \ref{tab:Directionality}.

Knowing $p_x$, $p_y$, and $p_z$, the azimuthal and zenith angles from the reactor can now be calculated:

\begin{equation}
  \begin{gathered}
    \tan \phi = \frac{p_y}{p_x} \\
    \tan \theta = \frac{p_z}{\sqrt{p_x^2 + p_y^2}} .
  \end{gathered}
  \label{eq:angle_tan}
\end{equation}

By taking the derivatives of Equation~\ref{eq:ratio} with the five uncorrelated, background-subtracted variables ($n_+,n_-,n_{0+},n_{0-},n_{0*}$), the uncertainty in $p_x$ and $p_y$ can be calculated. The uncertainty of the mean of the fit of the $z$ distribution gives us the uncertainty in $p_z$. These rectangular uncertainties are propagated through the coordinate transformations required to transform the Cartesian coordinates to spherical coordinates. This gives us the \num{1}$\sigma$ uncertainties in each angle. The total (background-subtracted) number of IBD candidates for the calculation of $p_x$, $p_y$ and $p_z$ are: \num{38307}, \num{37867}, and \num{48155} respectively. Note that the total IBD candidates reported here vary slightly for each direction. Without the additional selection requirements, the total number would be \num{61029} as in Reference \cite{PROSPECT_PRL}.

\begin{table}
  \setlength{\tabcolsep}{4pt} 
  \centering
  \begin{tabular}{| c | c | c | c |}
    \hline
    \textbf{}    & \textbf{$p_x$} (\unit{\milli \metre}) & \textbf{$p_y$} (\unit{\milli \metre}) & \textbf{$p_z$} (\unit{\milli \metre}) \\
    \hline
    Data         & \num{10.84 \pm 0.60}                  & \num{9.21 \pm 0.60}                   & \num{-1.87 \pm 0.39}                  \\
    \hline
    Data, biased & \num{8.70 \pm 0.37}                   & \num{7.31 \pm 0.37}                   & \num{-1.87 \pm 0.39}                  \\
    \hline
  \end{tabular}
  \caption{The values of $p_x$, $p_y$, and $p_z$ (average displacements in each direction) for the data with and without applying the modified method. "Data, biased" utilizes Equation \ref{eq:px} while "Data" uses Equation \ref{eq:ratio}. The uncertainties on the data values are statistical only.}
  \label{tab:Directionality}
\end{table}

\section{Verifications}

\subsection{BiPo}

To verify that the reconstruction of the antineutrino direction is unbiased, a cross-check is performed using intrinsic radioactivity. If there is any source of asymmetry in the methodology or detector topology, then it would affect every attempt to measure directionality. A convenient source of coincident events is the decay ${}^{214}$Bi $\rightarrow$ ${}^{214}$Po $\rightarrow$ ${}^{210}$Pb, part of the ${}^{222}$Rn decay chain.

The "BiPo" (bismuth polonium) decay is expected to be isotropic. ${}^{214}$Bi beta decays with a Q-value of \qty{3.28}{\mega \eV} \qty{99.98}{\percent} of the time. This beta decay is accompanied by one or more gamma rays \qty{81}{\percent} of the time. ${}^{214}$Po alpha decays with a kinetic energy of \qty{7.69}{\mega \eV} \qty{99.99}{\percent} of the time. The half life of ${}^{214}$Po is \qty{164}{\micro \second}. Both the BiPo decay and IBD are time coincidences between an electronic recoil event (positron/beta particle and gamma ray) and a diffuse event (neutron capture/alpha decay). Both occur in a similar time window of $\mathcal{O}$(\qty{100}{\micro \second}).

The event selection criteria for correlated BiPo events are the same used in reference \cite{PROSPECT_PRD}. The decay is seen at a rate of \qty{165}{\milli \hertz} or around \num{14000} per day in the PROSPECT-I detector.


Applying the same directionality technique to the BiPo decays seen in PROSPECT, any systematic biases that need to be considered in the directional reconstruction can be found. The distance between the prompt (beta/gamma) and delayed (alpha) events was calculated for both correlated and accidental BiPo candidates, then the accidentals were subtracted. Reactor Off data is used due to a  lower rate of gamma backgrounds. The same methods for calculating the $p_x$, $p_y$, $p_z$ values are applied here, so the effects for all three directions individually can be seen. Approximately \num{500000} BiPo events in each direction were utilized in this analysis. The results are in Table \ref{tab:bipo_result}.

\begin{table}
  \setlength{\tabcolsep}{7pt} 
  \centering
  \begin{tabular}{| c | c |}
    \hline
    \multicolumn{2}{|c|}{\textbf{BiPo Displacement Mean}}     \\
    \hline
    $p_x$ & \num{-0.25} $\pm$ \num{0.08} \unit{\milli \metre} \\
    \hline
    $p_y$ & \num{-0.39} $\pm$ \num{0.08} \unit{\milli \metre} \\
    \hline
    $p_z$ & \num{0.06} $\pm$ \num{0.06} \unit{\milli \metre}  \\
    \hline
  \end{tabular}
  \caption{The results for the average displacement in each direction for the BiPo events. Only the $z$ result agrees with our expected result of 0 mm within error. However, the $p_x$ and $p_y$ values are only a fraction of a mm away from zero.}
  \label{tab:bipo_result}
\end{table}

While $p_z$ is consistent with zero as expected, $p_x$ and $p_y$ show a slight bias on the sub-\unit{\milli \metre} scale. The source of this discrepancy is unknown. A similar discrepancy can be induced in simulations by unevenly dispersing BiPo events within the segments in $x$ and $y$. BiPo events were simulated in either only the top or bottom half of each segment in the detector. This caused a large bias of several millimeters, but since we have no ability to measure the $x$ or $y$ distributions of events within segments, it is unknown if a non-uniform BiPo distribution is the cause. Since it cannot be ruled out that the discrepancy also applies to IBD event-pairs, the BiPo mean and uncertainty were added in quadrature to the uncertainty of each component of $p$ for the IBDs as a systematic uncertainty. As the BiPo offset is smaller than the observed offset of neutrons from the IBD interaction point, as well as the statistical uncertainty on IBD event-pairs, the addition is a sub-dominant effect on total uncertainty.

\subsection{Simulations}
\label{subsec:sims}

To validate the methods described in Section \ref{sec:methods}, simulations were generated using PG4, a GEANT-4 \cite{GEANT4} based simulation of the PROSPECT-I detector. The IBD modeling in PG4 is based on the Vogel and Beacom model \cite{Vogel}. The modeled detector response is adjusted to reproduce the observed PROSPECT-I data for a wide variety of radioactive calibration sources and intrinsic background energy depositions. PG4 utilizes a cylindrical neutrino source whose center is placed at the location of the \HFIR\ core, determined from detailed drawings of the \HFIR\ facility as well as the PROSPECT-I detector's location within it.  The measurements are conservatively estimated to be accurate to \qty{10}{\centi \metre} in $x$, $y$, and $z$. With the origin placed at the center of the detector, the location of the center of the \HFIR\ core is: $x$ = \qty{-5.97}{\metre}, $y$ = \qty{-5.09}{\metre}, and $z$ = \qty{1.19}{\metre}. The estimated "true" antineutrino direction, from the center of the reactor core to the average prompt location within the PROSPECT-I detector, is: $\phi$ = \qty{40.8 \pm 0.7}{\degree} and $\theta$ = \qty{98.6 \pm 0.4}{\degree}. The average prompt location was calculated using a weighted average of simulation events over the five periods of PROSPECT data.

\begin{figure}
  \centering
  \includegraphics[width = 0.42\textwidth]{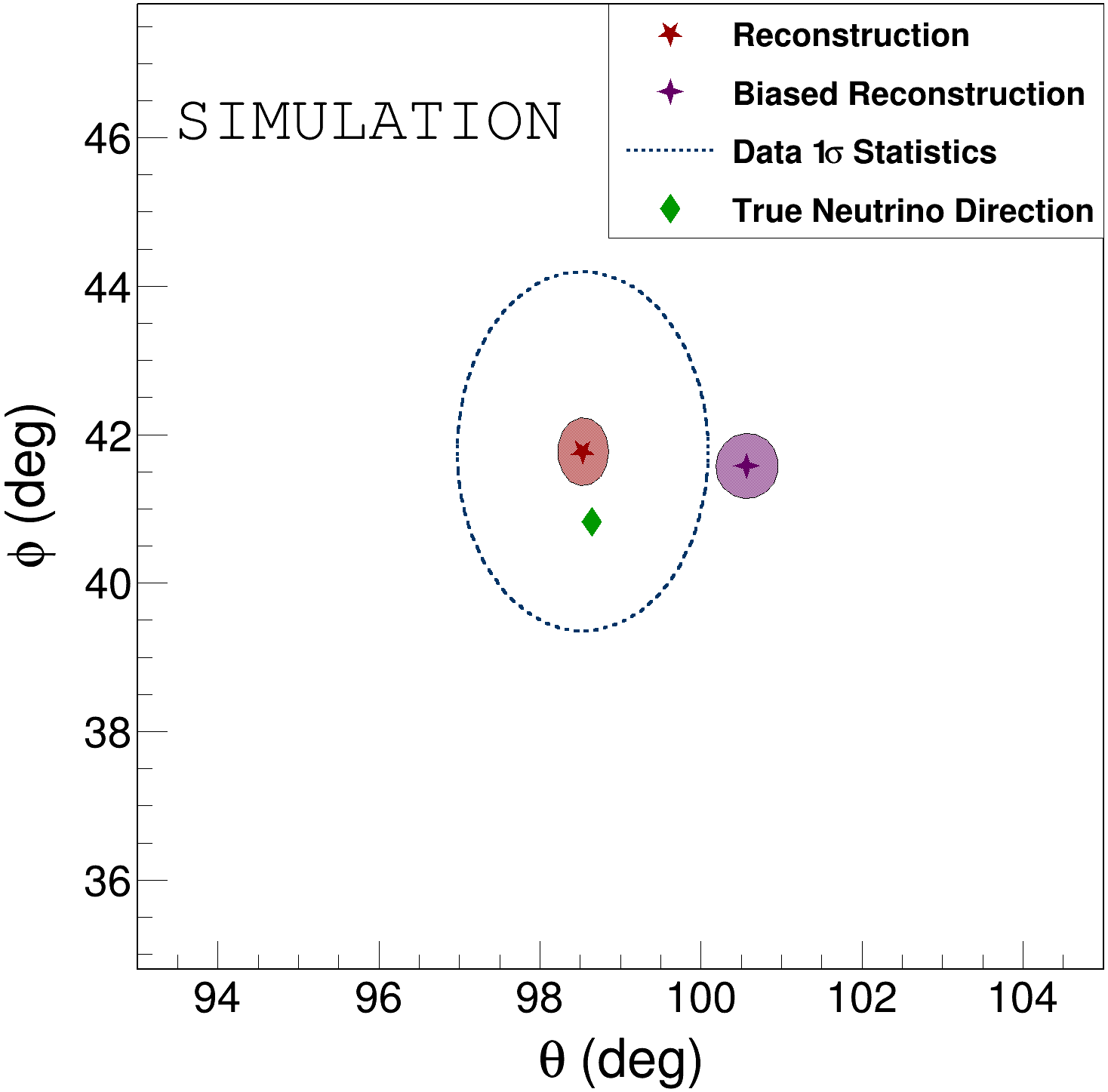}
  \caption{Reconstructed average antineutrino direction for the simulations, with \num{1}$\sigma$ uncertainty ellipses. "Biased Reconstruction" utilizes Equation \ref{eq:px} while "Reconstruction" uses Equation \ref{eq:ratio}. The statistical uncertainty for the IBD dataset is also included as a dashed ellipse, centered on the simulation direction. The simulation without bias mitigation shows a clear systematic shift. A smaller systematic shift remains after bias mitigation.}
  \label{fig:sim_plot}
\end{figure}

Simulations were produced with the same segment configurations as each of the five periods. The combined simulation should thus experience the same effects from the inoperative segments. The simulations were generated to give ten times the number of signal events as the data after the selection process. The reconstructed average antineutrino direction from the simulation, using the methods described in Section \ref{sec:methods}, is shown in Figure~\ref{fig:sim_plot}. To illustrate the impact of the unbiasing method, the biased simulation result using Equation \ref{eq:px} instead of Equation \ref{eq:ratio} for $p_x$ and $p_y$ is also plotted in Figure \ref{fig:sim_plot}. The dashed ellipse in Figure \ref{fig:sim_plot} showcases the statistical uncertainty of the IBD dataset, centered on the best fit marker for the simulation result. As expected, the biased method (using Equation \ref{eq:px}) shows a significant disagreement with the true neutrino direction. The offset in $\phi$ is \ang{0.2} while $\theta$ is offset by \ang{2.4}.

After unbiasing, an offset of \ang{0.9} in $\phi$ and \ang{0.1} in $\theta$ remains between the reconstructed and true directions in simulation, due to unaccounted for detector effects. In particular, the \ang{5.5} tilt of the segments is estimated to result in an offset of roughly this size and direction. This offset is corrected for in the data, with a systematic uncertainty of \qty{100}{\percent} applied to the correction.

\section{Results}

\begin{table}
  \setlength{\tabcolsep}{4pt} 
  \centering
  \begin{tabular}{| c | c | c | c |}
    \hline
    \textbf{}          & $p_x$ (\unit{\milli \metre}) & $p_y$ (\unit{\milli \metre}) & $p_z$ (\unit{\milli \metre}) \\
    \hline
    Data               & \num{10.84 \pm 0.65}         & \num{9.21 \pm 0.72}          & \num{-1.87 \pm 0.40}         \\
    \hline
    Data, biased       & \num{8.70 \pm 0.45}          & \num{7.31 \pm 0.54}          & \num{-1.87 \pm 0.40}         \\
    \hline
    Simulation         & \num{10.90 \pm 0.12}         & \num{9.74 \pm 0.12}          & \num{-2.20 \pm 0.08}         \\
    \hline
    Simulation, biased & \num{8.76 \pm 0.09}          & \num{7.77 \pm 0.09}          & \num{-2.20 \pm 0.08}         \\
    \hline
  \end{tabular}
  \caption{The values of $p_x$, $p_y$, and $p_z$ (average displacements in each direction) for the data and simulations. The uncertainties on the data values are statistical and systematic uncertainties combined in quadrature. The simulation and real data both employ the same analysis method. The biased simulation utilizes Equation \ref{eq:px} for calculating $p_x$ and $p_y$. We note that the addition of systematic uncertainties only causes a slight increase from the statistical uncertainties shown in Table \ref{tab:Directionality}.}
  \label{tab:Directionality_Systematics}
\end{table}

\begin{figure}
  \centering
  \includegraphics[width = 0.42\textwidth]{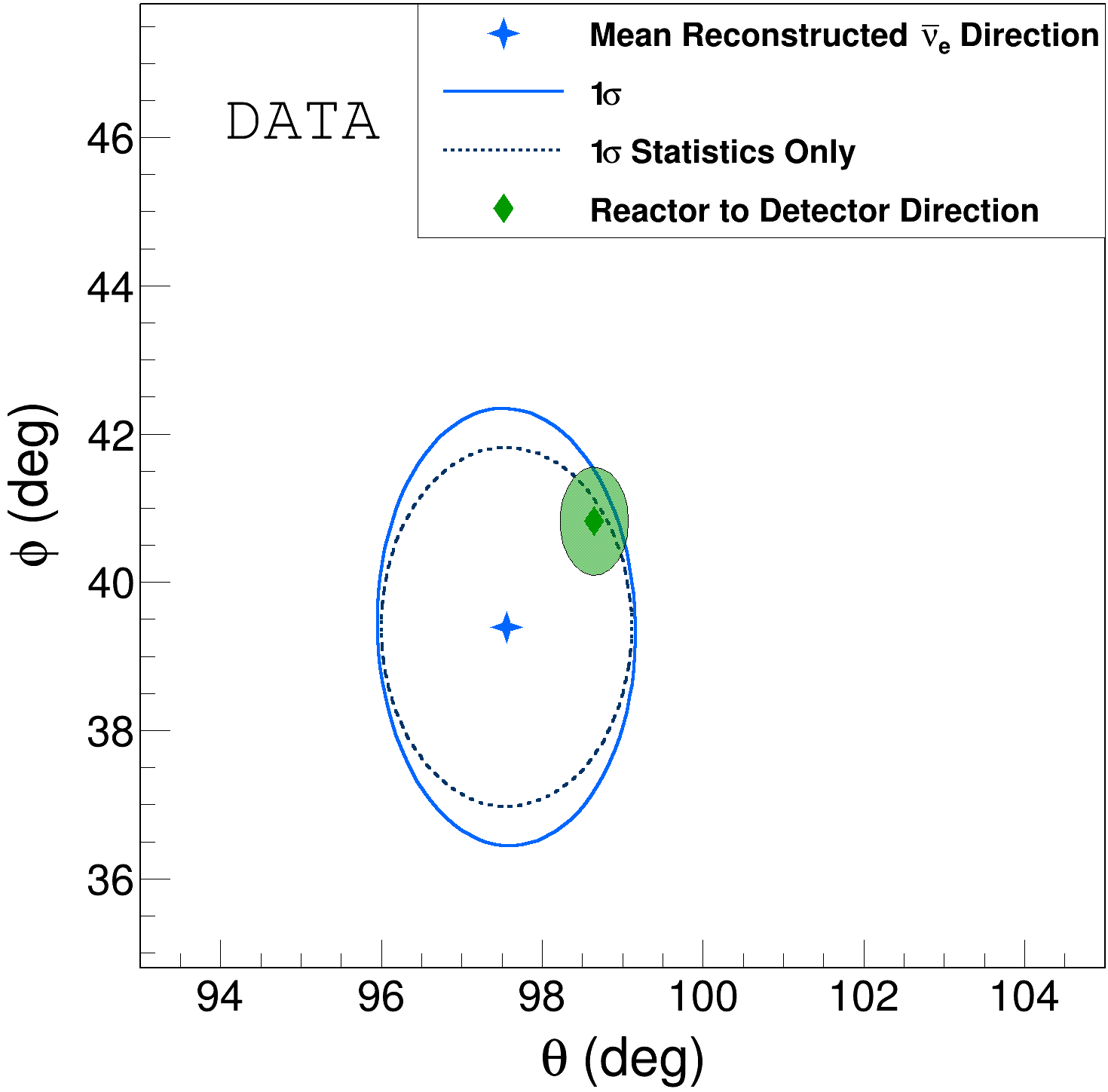}
  \caption{Reconstructed average antineutrino direction, with \num{1}$\sigma$ uncertainty ellipses. The addition of systematics causes a visible but slight decrease in precision, as the uncertainty is statistically dominated. The "Reactor to Detector Direction" point illustrates the angles from the center of the reactor to the average prompt location in the detector. The green ellipse around the point is derived from the the \qty{10}{\centi \meter} $x$, $y$ and $z$ uncertainties in the relative locations of the detector and the reactor core from the HFIR diagrams.}
  \label{fig:final_plot}
\end{figure}

The measured values for $p_x$, $p_y$, and $p_z$ for data and simulation are shown in Table \ref{tab:Directionality_Systematics}. The calculated azimuthal and zenith angles from the data are: $\phi$ = \qty{39.4 \pm 2.4}{\degree} (stat.) \qty{\pm 1.9}{\degree} (syst.) and $\theta$ = \qty{97.6 \pm 1.5}{\degree} (stat.) \qty{\pm  0.5}{\degree} (syst.). These results are plotted in Figure \ref{fig:final_plot}.

We note that the simulation and data results agree with the true neutrino direction and estimated reactor to detector direction shown in the figures, verifying PROSPECT's ability to reconstruct the direction of the reactor core. This measurement is dominated by statistics, and would show greater precision with a higher number of events. The systematic uncertainties are greater for $\phi$ than $\theta$ for two reasons. The first is that the offset correction applied to $\phi$ is greater than $\theta$. Second is that $\phi$ is only dependent on $p_x$ and $p_y$, and the magnitude of the means for BiPo are higher for $p_x$ and $p_y$ than $p_z$.

For comparison, the most precise measurements previously conducted by the CHOOZ experiment can be considered \cite{CHOOZ}. CHOOZ achieved a \num{2}-dimensional statistical uncertainty of \ang{18} with roughly \num{2700} IBD event-pairs. This uncertainty is defined as the half-angle of a cone centered on the true direction, with a \qty{68}{\percent} probability of containing the measured direction. Double Chooz improved this result to \ang{9.4} with \num{8249} events \cite{Caden}. As PROSPECT's uncertainties are asymmetric, the ellipse in Figure \ref{fig:final_plot} can be utilized. The statistical ellipse subtends the same solid angle as a cone with a half-angle of \ang{2.9}.

This showcases an improvement of a factor of \num{3} compared to Double Chooz and \num{6} compared to CHOOZ. Approximately a factor of two improvement results from PROSPECT’s improved position resolution. The remainder is largely due to increased statistics. The improved position resolution of PROSPECT-I can be seen in the reconstructed distance between prompt and delayed events for a typical event pair, a quantity broadened by detector effects. In PROSPECT-I, the typical (mean)  resolution is \qty{11}{\centi \metre}, whereas for CHOOZ it is approximately \qty{25}{\centi \metre} \cite{CHOOZ}. These can be compared to the true average spread between positron and neutron capture, which in PROSPECT simulations is \qty{6}{\centi \metre} and is expected to be similar for CHOOZ. The improvement is in large part due to the ${}^6$Li neutron capture target, and directly translates to improvement in the pointing accuracy. Effective counts provide the best comparison to CHOOZ and Double Chooz statistics, as roughly background-free equivalent statistics. Effective counts in PROSPECT are defined as:

\begin{equation}
  N_{\texttt{eff}} = \frac{N_{\texttt{IBD}}^2}{\sigma_{\texttt{IBD}}^2} ,
\end{equation}

where $N_{\texttt{eff}}$ is effective counts, $N_{\texttt{IBD}}$ is background-subtracted IBD counts, and $\sigma_{\texttt{IBD}}$ is the statistical uncertainty, including uncertainty from background subtraction.The effective counts for $p_x$, $p_y$, and $p_z$ are: \num{22200}, \num{22058}, and \num{28939} respectively. PROSPECT-I's dataset has $\sim$\num{10}$\times$ more effective events than CHOOZ and $\sim$\num{3}$\times$ more than Double Chooz. The pointing accuracy is expected to be proportional to $\sqrt{N_{\texttt{eff}}}$.

\section{Conclusions}

The PROSPECT experiment has conducted the first antineutrino direction measurement using a {}$^6$Li doped liquid scintillator. We report the azimuthal and zenith angles as $\phi$ = \qty{39.4 \pm 2.9}{\degree} (stat. + sys.) and $\theta$ = \qty{97.6 \pm 1.6}{\degree} (stat. + sys.). This direction agrees with the known direction from the reactor to the detector. The angles were reconstructed with approximately \num{48000} IBD event-pairs collected over \num{95.62} Reactor On days. The improvements afforded by PROSPECT-I's segmentation, {}$^6$Li doping, and higher statistics allow us to provide the most precise measurement of reactor antineutrino direction so far.

These techniques provide evidence that larger experiments utilizing $^6$Li-doped liquid scintillators at kiloton mass scales will be able to conduct precision measurements with the advantage of notably larger statistics. The PROSPECT-I simulations also show that higher statistics would significantly reduce the uncertainty.

Future work could study the ability of segmented detectors like PROSPECT-I to resolve multiple antineutrino sources considering factors like segment pitch, source-detector orientation, and event statistics. The results also provide confidence in the event reconstruction capabilities and analysis methods utilized by the PROSPECT experiment. It is important to note that these results showcase the ability to determine the angular direction of a point source, not an angular resolution.

\section{Acknowledgements}
This material is based upon work supported by the following sources: US Department of Energy (DOE) Office of Science, Office of High Energy Physics under Award No. DE-SC0016357 and DE-SC0017660 to Yale University, under Award No. DE-SC0017815 to Drexel University, under Award No. DE-SC0008347 to Illinois Institute of Technology, under Award No. DE-SC0016060 to Temple University, under Award No. DE-SC0010504 to University of Hawaii, under Contract No. DE-SC0012704 to Brookhaven National Laboratory, and under Work Proposal Number  SCW1504 to Lawrence Livermore National Laboratory. This work was performed under the auspices of the U.S. Department of Energy by Lawrence Livermore National Laboratory under Contract DE-AC52-07NA27344 and by Oak Ridge National Laboratory under Contract DE-AC05-00OR22725. Additional funding for the experiment was provided by the Heising-Simons Foundation under Award No. \#2016-117 to Yale University.

We further acknowledge support from Yale University, the Illinois Institute of Technology, Temple University,  University of Hawaii, Brookhaven National Laboratory, the Lawrence Livermore National Laboratory LDRD program, the National Institute of Standards and Technology, and Oak Ridge National Laboratory. We gratefully acknowledge the support and hospitality of the High Flux Isotope Reactor and Oak Ridge National Laboratory, managed by UT-Battelle for the U.S. Department of Energy.

\bibliography{references}

\end{document}

%% file: AuthorList2022.tex
\affiliation{Department of Physics, Boston University, Boston, MA, USA} \vspace{-0.4\baselineskip}
\affiliation{Brookhaven National Laboratory, Upton, NY, USA} \vspace{-0.4\baselineskip}
\affiliation{Department of Physics, Drexel University, Philadelphia, PA, USA} \vspace{-0.4\baselineskip}
\affiliation{George W.\,Woodruff School of Mechanical Engineering, Georgia Institute of Technology, Atlanta, GA, USA} \vspace{-0.4\baselineskip}
\affiliation{Department of Physics and Astronomy, University of Hawaii, Honolulu, HI, USA} \vspace{-0.4\baselineskip}
\affiliation{Department of Physics, Illinois Institute of Technology, Chicago, IL, US} \vspace{-0.4\baselineskip}
\affiliation{Nuclear and Chemical Sciences Division, Lawrence Livermore National Laboratory, Livermore, CA, USA} \vspace{-0.4\baselineskip}
\affiliation{Department of Physics, Le Moyne College, Syracuse, NY, USA} \vspace{-0.4\baselineskip}
\affiliation{National Institute of Standards and Technology, Gaithersburg, MD, USA} \vspace{-0.4\baselineskip}
\affiliation{High Flux Isotope Reactor, Oak Ridge National Laboratory, Oak Ridge, TN, USA} \vspace{-0.4\baselineskip}
\affiliation{Physics Division, Oak Ridge National Laboratory, Oak Ridge, TN, USA} \vspace{-0.4\baselineskip}
\affiliation{Department of Physics, Susquehanna University, Selinsgrove, PA, USA} \vspace{-0.4\baselineskip}
\affiliation{Department of Physics, Temple University, Philadelphia, PA, USA} \vspace{-0.4\baselineskip}
\affiliation{Department of Physics and Astronomy, University of Tennessee, Knoxville, TN, USA} \vspace{-0.4\baselineskip}
\affiliation{Department of Physics, United States Naval Academy, Annapolis, MD, USA} \vspace{-0.4\baselineskip}
\affiliation{Department of Physics, University of Wisconsin, Madison, WI, USA} \vspace{-0.4\baselineskip}
\affiliation{Wright Laboratory, Department of Physics, Yale University, New Haven, CT, USA} \vspace{-0.4\baselineskip}

\author{M.~Andriamirado$^{6}$,
    A.~B.~Balantekin$^{16}$,
    C.~D.~Bass$^{8}$,
    O. Benevides Rodrigues$^{6}$,
    E.~P.~Bernard$^{7}$,
    N.~S.~Bowden$^{7}$,
    C.~D.~Bryan$^{10}$,
    R.~Carr$^{ 15}$,
    T.~Classen$^{7}$,
    A.~J.~Conant$^{10}$,
    G.~Deichert$^{10}$,
    M.~J.~Dolinski$^{3}$,
    A.~Erickson$^{4}$,
    A.~Galindo-Uribari$^{11,14}$,
    S.~Gokhale$^{2}$,
    C.~Grant$^{1}$,
    S.~Hans$^{2}$,
    A.~B.~Hansell$^{12}$,
    K.~M.~Heeger$^{17}$,
    B.~Heffron$^{11,14}$,
    D.~E.~Jaffe$^{2}$,
    S.~Jayakumar$^{3}$,
    D.~C.~Jones$^{13}$,
    J.~Koblanski$^{5}$,
    P.~Kunkle$^{1}$,
    C.~E.~Lane$^{3}$,
    B.~R.~Littlejohn$^{6}$,
    A. Lozano Sanchez$^{3}$,
    X.~Lu$^{11,14}$,
    J.~Maricic$^{5}$,
    M.~P.~Mendenhall$^{7}$,
    A.~M.~Meyer$^{5}$,
    R.~Milincic$^{5}$,
    P.~E.~Mueller$^{11}$,
    H.~P.~Mumm$^{9}$,
    J.~Napolitano$^{13}$,
    C. J. Nave$^{3}$,
    R.~Neilson$^{3}$,
    M. Oueslati$^{3}$,
    C.~Roca$^{7}$,
    R.~Rosero$^{2}$,
    P.~T.~Surukuchi$^{17}$,
    F.~Sutanto$^{7}$,
    D.~Venegas-Vargas$^{11,14}$,
    P.~B.~Weatherly$^{3}$,
    J.~Wilhelmi$^{17}$,
    M.~Yeh$^{2}$,
    C.~Zhang$^{2}$ and
    X.~Zhang$^{7}$ \\
}
